\documentclass[conference]{IEEEtran}
\IEEEoverridecommandlockouts
\usepackage{cite}
\usepackage{amsmath,amssymb,amsfonts}
\usepackage{algorithmic}
\usepackage{graphicx}
\usepackage{textcomp}
\usepackage{xcolor}
\usepackage{hyperref}
\usepackage{tcolorbox}
\usepackage{ragged2e}
\usepackage{graphicx}
\usepackage{amsmath}
\usepackage{algorithm}
\usepackage{algorithmic}
\usepackage{adjustbox}
\usepackage{comment}
\usepackage{soul}
\usepackage{microtype} 
\usepackage{flushend}

\usepackage[utf8]{inputenc}
\usepackage[T1]{fontenc}
\usepackage{microtype}

\def\BibTeX{{\rm B\kern-.05em{\sc i\kern-.025em b}\kern-.08em
    T\kern-.1667em\lower.7ex\hbox{E}\kern-.125emX}}
\begin{document}

\title{Towards a Probabilistic Framework for Analyzing and Improving LLM-Enabled Software

\thanks{Partially supported by UBACyT Grants 20020220300079BA and 20020190100126BA, CONICET PIP 11220200100084CO, and ANPCyT PICT-2021-I-A-00755, A-1-2022-1-173516 IDRC-ANII, SWPERFI UFAM-MOTOROLA RD\&I Project ``Técnicas de Inteligência Artificial para Análise e Otimização de Desempenho de Software'', and Amazon Research Award -- Fall 2023 on Automated Reasoning.}
}


%


\author{\IEEEauthorblockN{Juan Manuel Baldonado}
\IEEEauthorblockA{\textit{ICC UBA/CONICET and DC, FCEN} \\
\textit{Universidad de Buenos Aires}\\
Buenos Aires, Argentina \\
juanmanuelbaldonado@gmail.com}
\and
\IEEEauthorblockN{Flavia Bonomo-Braberman}
\IEEEauthorblockA{\textit{ICC UBA/CONICET and DC, FCEN} \\
\textit{Universidad de Buenos Aires}\\
Buenos Aires, Argentina \\
fbonomo@dc.uba.ar}
\and
\IEEEauthorblockN{V\'{\i}ctor A. Braberman}
\IEEEauthorblockA{\textit{ICC UBA/CONICET and DC, FCEN} \\
\textit{Universidad de Buenos Aires}\\
Buenos Aires, Argentina \\
vbraber@dc.uba.ar}
}

\maketitle

\begin{abstract}
Ensuring the reliability and verifiability of large language model (LLM)-enabled systems remains a significant challenge in software engineering. We propose a probabilistic framework for systematically analyzing and improving these systems by modeling and refining distributions over clusters of semantically equivalent outputs. This framework facilitates the evaluation and iterative improvement of Transference Models--key software components that utilize LLMs to transform inputs into outputs for downstream tasks. To illustrate its utility, we apply the framework to the autoformalization problem, where natural language documentation is transformed into formal program specifications. Our case illustrates how distribution-aware analysis enables the identification of weaknesses and guides focused alignment improvements, resulting in more reliable and interpretable outputs. This principled approach offers a foundation for addressing critical challenges in the development of robust LLM-enabled systems.
\end{abstract}

\begin{IEEEkeywords}
LLMs, prompt engineering,  autoformalization
\end{IEEEkeywords}

\section{Introduction}
In recent years, Large Language Models (LLMs), such as GPT-4~\cite{openai2024gpt4technicalreport} and Gemini~\cite{geminiteam2024geminifamilyhighlycapable}, have demonstrated remarkable capabilities across diverse applications. These successes are largely attributed to their instruction-following abilities and in-context learning. By conditioning on task-specific instructions (zero-shot) or a small set of examples (few-shot), LLMs have been shown to perform a wide array of tasks effectively~\cite{Few-Shot-Brown20}. This adaptability has led to a proliferation of applications leveraging LLMs to elicit various downstream tasks via prompting.
The integration of LLMs into software systems is a rapidly growing trend, but it presents numerous challenges~\cite{hassan2024rethinking}. Particularly, phenomena such as hallucination~\cite{DBLP:journals/csur/JiLFYSXIBMF23} impact what can be guaranteed about such applications. 
One basic aspect to disciplined engineering of complex systems is how to understand and analyze their behavior. While NLP research acknowledges 
and leverages at some extent the underlying predictive model on next token distribution (e.g.,~\cite{DBLP:journals/nature/FarquharKKG24,Self-consist}), research papers on software engineering of LLM-enabled applications typically pinpoint that aspect just as a troublesome source of non-determinism  (e.g., flaky tests~\cite{hassan2024rethinking}).
Instead, we believe that a disciplined engineering of reliable LLM-enabled applications should be based on both the identification of constituent  ``transference models'' (TMs) --LLM-powered software components transforming inputs into outputs--, and the modeling and analysis of the underlying distribution over classes that comprise (transference)-equivalent outputs (i.e., ``meaning classes''~\cite{DBLP:journals/nature/FarquharKKG24}). 
In particular, we propose an evaluation and refinement framework based on the above insights and also on the belief that it is feasible and worthwhile to characterize in which extent and how distribution over meaning-classes satisfies or deviates from expected distribution characteristics. More concretely, we hypothesize that, for inputs in which TM yields concentrated distributions on incorrect meaning-classes (adversarial cases), it is possible to verbalize into task-specific terms the kind of misalignment, and that constitutes a valuable input for re-engineering transference models (e.g., sub-task decomposition, replacement of underlying LLM, or other targeted adjustments informed by such analyses).

To illustrate these ideas, we present a preliminary case study addressing the problem of autoformalization: translating docstrings that specify the intention of a code snippet into formal pre- and post-conditions (e.g., assertions written in Dafny~\cite{dafny}). This downstream task, which translates natural language into formal languages, is particularly useful in software testing and verification~\cite{Clover}. Our study demonstrates how probabilistic analysis and meaning-class transformations can lead to focused alignment improvements, offering a foundation for principled advancements in LLM-enabled systems.

\section{Preliminaries}
\subsection{Large Language Models}
These predictive models are parameterized functions, whose parameter sets are often initially found by optimizing an objective or loss function for next-token prediction with respect to a training corpus. 
Thus, at its core, an LLM yields a probability distribution over the set of tokens or vocabulary (i.e.,  $P(t_k|x)$, the probability of the $k$-th token being $t_k$ given the context $x$).  
LLMs are typically used in its generative role by performing some particular decoding approach using the next-token predictive model~\cite{10.1162/tacl_a_00502}. 
In general, LLM-enabled software leverages them by injecting prompts that condition continuations/responses (i.e., LLM predicts next-tokens conditioned by the prompt), thus hopefully eliciting the desired downstream tasks.
Task-agnostic prompting strategies are frequently used (e.g.~\cite{CoT-Wei22,RAG2020Patrick}).

\subsection{Autoformalization}
As indicated, our illustration is based on the problem of autoformalization in software verification. This is the transformation from natural language descriptions (docstrings, typically available in development processes) into some formally correct and automatically verifiable format \cite{AutoformalizationSzegedy,Clover} which enables automated reasoning on program's correctness. 
In our illustration, we choose Dafny~\cite{dafny} as the target formal language and particularly the pre- and post-condition sections. 
\begin{figure}[h]
    \centering
    \includegraphics[width=\columnwidth]{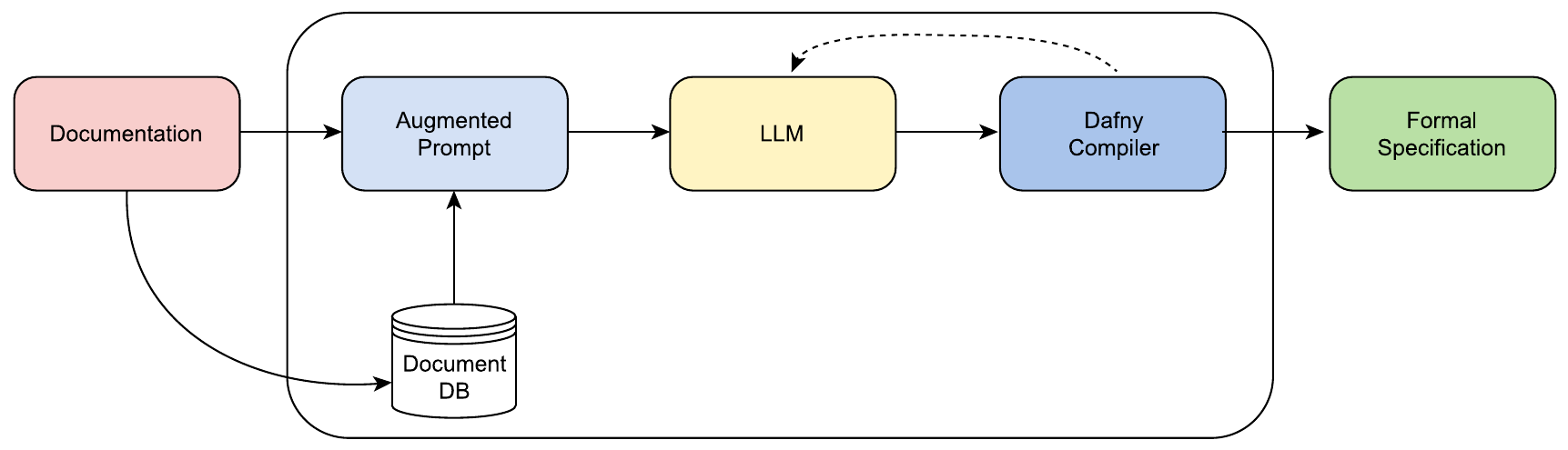}
    \caption{Architecture of the illustrated baseline transference model}
    \label{fig:arch}
\end{figure}

\subsection{Transference Models} 

Following~\cite{Clover}, we define ``Transference Models'' (TMs) as software components that use (typically, prompt-modulated) LLMs to elicit downstream tasks as the key (but not necessarily unique) means for transforming some input data into some output data. 
Those transformations have potentially stochastic nature given the underlying predictive model and (sampling-based) decoding strategy of the LLMs. Thus, TMs behavior can be formulated as a function of input-output pairs into reals, that is, $T: \mathcal{I} \times \mathcal{O} \rightarrow \mathbb{R}$, where $T(i, o)$ denotes the probability that $i \in \mathcal{I}$ is transferred to $o \in \mathcal{O}$, and for $i \in \mathcal{I}$, $T(i,\cdot)$ is a probability distribution over~$\mathcal{O}$.
As an example, in autoformalization proposals, typically, the TM is a stochastic process that links natural language with annotations as the expected type of input-output pairs (e.g.,~\cite{Clover,braberman2024tasks}). 
Figure~\ref{fig:arch} shows the architecture of the TM developed for illustration. Similarly to Clover's ``doc2anno'' TM~\cite{Clover}, it elicits a formalization task and, if Dafny compiler detects that the generated annotation is syntactically incorrect, it instructs a corrective version of the formalization task by including compiler's feedback a bounded number of times. We also include RAG-like (few-shot) prompt generation~\cite{RAG2020Patrick} to get annotations for likely similar problems by querying a vector database.
\subsection{Distributions over Semantic Domains}
Given that Language Models are density estimation functions that assign probability to every possible string,
we concur with others (e.g.~\cite{DBLP:conf/emnlp/HoltzmanWSCZ21,DBLP:journals/nature/FarquharKKG24}) that probability should be deemed assigned to concepts, not strings: there are often many (or even infinite) strings that represent a given idea equally well (e.g., in the autoformalization problem, logically equivalent annotations) and share the same meaning when embedded into the abstract domain $\mathcal{O}$. This is, in theory, the result of summing up the probabilities assigned to strings in each of the equivalence classes. 
\subsection{Computing Empirical Distributions of TMs}
Given the potential richness of  TMs and decoding strategies  plus the fact next-token probabilities are not always accessible,  for a given input, we approximate the TM probability distribution by the empirical categorical distribution on meaning-classes that are identified and built on-the-fly by clustering generated outputs (e.g., by checking annotations to be SMT-equivalent). 
More concretely, re-execution of code of the transference model with the same input 
triggers the (many times stochastic) decoding adapters~\cite{10.1162/tacl_a_00502}
of the involved LLMs and thus the entire TM behaves as a black box stochastic process (whose behavior depends on hyperparameters and settings of those adapters like temperature, etc.). Yielded outputs are then clustered according to the equivalence relation and empirical distribution on classes is computed (see similar discussion in~\cite{DBLP:journals/nature/FarquharKKG24}). 
\section{The Framework}
\subsection{Assumptions, Hypothesis and Rationale}
Firstly, we assume that for most LLM-enabled applications, it is possible to understand their LLMs interactions as part of the implementation of one or more TMs.  TMs might be implemented straightforwardly by zero-shot instruction prompting that includes some representation of the input. However, TMs could also be much more sophisticated or even generated by automated prompt generation frameworks like~\cite{DSPY}. As illustrated above, they could include statically or dynamically orchestrated external tools (e.g., some feedback signal for corrective tasks), general reactive LLM-triggered external computation~\cite{ReAct}, the chaining of a series of lower-level transference models~\cite{DSPY}, etc. 
TMs are the  component under analysis in our approach. Thus the second assumption is that those input/output transformations  can be recovered and described  as --human understandable-- downstream tasks either by developers/testers (or LLMs, in the future) (e.g.~\cite{braberman2024tasks}). This understanding should include the ability to assess and equate tasks results into conceptual classes.   
Our main hypothesis is that to gain insight on the potential behavior of an LLM-based TM (e.g., when testing it, when evaluating it, when validating it, when engineering it, when over-sighting it, etc.) it is key to embrace the predictive nature of underlying model and thus approximate and characterize the yielded underlying probability distribution of TMs over concepts\footnote{We also hypothesize that consistent probability assignments for task's equivalent results is likely an emergent behavior of LLMs.}.   
More concretely, we believe that concentrated and misaligned cases are the ones to be detected and reduced during development time. 
The rationale is manyfold: on the one hand, concentrated and misaligned cases pinpoint to (adversarial) inputs that lead to systematic misalignment  in the prompt-modulated predictive model while they could potentially deceive certainty-based hallucination detection techniques~\cite{DBLP:journals/nature/FarquharKKG24} (i.e., false negatives that would need to be further identified or mitigated somehow in a safe way).
Also, it has been shown that assuming TMs to be aligned and concentrated with high probability may provide error-bounded performance guarantees (e.g.~\cite{Clover}). 
Thus, improvement is defined to be akin refinement or behavioral subtyping used in  program verification settings (e.g.~\cite{DBLP:journals/toplas/LiskovW94}): replacing a TM with an improved one will mean less false positives for a certainty-based hallucination guardrail, if deployed, and a reduced likelihood of error. We also require (almost-sure) non-regression on aligned cases.  

On the other hand, we also hypothesize that, when a meaning class that gets the most of the probability mass does not match TM's expected result, 
the nature of failure can be, in general, stated in terms of mistakes  humans have already studied and categorized in the corresponding problem domain and could lead to hints regarding improvement. 
That is, we seek characterizing in \textit{task-specific} terms the nature of {\it failures}. 
For instance, for the autoformalization problem studied, ``too weak/strong pre/post-conditions'' 
are well-known concepts of formalization mistakes ontology. As we will show later, they are also typically the perfect fit for characterizing TM's failures.   
\subsection{Defining Improvement}
For a given input, TM's distribution over meaning classes is \textit{aligned} when the class with the largest assigned probability is a correct one, and \textit{misaligned} otherwise. 
\textit{Concentration} is used here as a practical proxy of certainty on generated concepts (akin semantic entropy of the empirical distribution), and it is defined here as a distribution characterized by the winning meaning class having probability mass greater than or equal to the sum of the rest of the masses. Other definitions are possible (e.g.,~\cite{Clover}).
Given two alternative $t,t'$ TMs, we say that TM $t'$  pointwise \textit{improves}  over $t$ for a test set when $t'$ extends the set of inputs which feature aligned and concentrated distributions and reduces  the set of inputs that feature concentrated and misaligned distributions. The generalized and computable definition of almost-sure improvement for the entire domain is possible by assuming some probability distribution $D$ over the domain of inputs (or a set of possible distributions). Although its definition and study is out of scope, it can be informally stated as a)~probability of an input sampled according $D$ to be related by $t'$ to a concentrated and aligned distribution is greater than sampling from $D$  an input with such characteristics in $t$, b)~with probability close to one, an input drawn from $D$ that has a concentrated and aligned distribution in $t$ also has an aligned and concentrated distribution in $t'$, and c)~probability of an input sampled according $D$ to be related by $t'$ to a concentrated and misaligned distribution is less or equal than sampling from $D$  an input with such characteristics in $t$. 

\section{Illustrative results on Autoformalization}

Next we show an anecdotal illustration in line with the stated hypothesis (no claim of generalization).  
We use an existing dataset of Dafny programs; the CloverBench dataset consists of 62 small hand-written example programs similar to those found in standard computer science textbooks~\cite{Clover}. Each program in the dataset consists of a single method and includes a Dafny implementation, annotations specifying pre-conditions and post-conditions, and a docstring that describes program's intention. We extract from those programs docstrings and method signatures and we use pre- and post-conditions as the groundtruth to assess alignment of the most probable meaning class. We get the empirical approximation by re-execution of the transference model (30 times in this case) and clustering with the assistance of a SMT solver. For the sake of simplicity, we treat each syntactically-invalid result as its own class, rather than assigning probability mass to the edit-distance closest valid class. Compared to collapsing all invalid results into a single class, it may lead to less concentration in verdicts, especially when the model assigns significant probability to diverse invalid outputs.
We opted for the \textit{Gemini~1.5 Flash} model with a context of 1000 tokens.  
Hyperparameters were the default ones for that LLM:  $top_k$ with $k=40$, $top_p$ with $p=0.95$ and temperature $0.7$. This means we work with a generative-process distribution that is more skewed (and tail-truncated) than the underlying prompt-modulated next-token prediction model distribution. Yet, this illustrates a plausible generative-process distribution of the LLM running at the core of the TM whose stochastic behavior one wants to approximate under such decoding settings.  More details and data can be found in~\cite{repo-juan}.

\begin{table}[h!]
   \caption{Alignment and Concentration of Distributions}
    \label{tab:alignment_concentration_exp1}
\centering
   \resizebox{\columnwidth}{!}{%
    \begin{tabular}{|c|c|c|}
        \hline
        \textbf{Alignment/Concentration} & \textbf{Concentrated} & \textbf{Not Concentrated} \\
        \hline
        Aligned & 46 & 3 \\
        \hline
        Correct class gen. but misaligned distr. & 2 & 3 \\
        \hline
        Correct class not generated & 4 & 4 \\
        \hline
    \end{tabular}}
 \end{table}
    In Table \ref{tab:alignment_concentration_exp1}, we show a breakdown of the results obtained in terms of alignment and concentration. For a given input (e.g., docstring + signature), we say that a meaning class (i.e., set of equivalent pre/post specifications) is \textit{correct}  when it is made up of specs equivalent to groundtruth specification.
Firstly, 54 out of the 62 inputs, the empirical distribution features a non-zero probability meaning-class which is the correct one (although not necessarily the winning one). For those 54 inputs, in 49 inputs the obtained empirical distribution is actually aligned. Moreover, in 46 cases, the distribution is \textit{both concentrated and aligned}: a sunny day scenario. It is worth noting that the baseline TM using zero-temp (almost) deterministic single-output regime in the underlying LLM gets 47 aligned results. However, TM's distributions on concepts for some of those inputs are either misaligned or they are high-entropy distributions. Thus, zero-temp disguises potential issues of the underlying model's probability assignment. On the other hand, there are also inputs in which the zero-temp utterance is incorrect but TM's distribution is actually aligned and concentrated. 
A similar phenomenon has been observed for close-ended selection tasks~\cite{DBLP:conf/emnlp/HoltzmanWSCZ21}.
We argue that for reliability and safety concerns one should focus on understanding and reducing cases in which misalignment concurs with high concentration. 
In fact, Table~\ref{tab:error_breakdown_exp1} yields the in-depth analysis of winning meaning-class in the misaligned  cases. As mentioned earlier, how the winning-class differs from the groundtruth can be many times defined naturally in task-specific terms. In fact, 11 out of 14 misalignments were  easily matchable to well-known bug types in formalization. 
When analyzing reasons for misalignment, weak post-condition is the more frequent characteristic of winning meaning-classes. One of the 2 cases in ``debugging'' focus (misaligned and concentrated) is \textit{LinearSearch}. It is supposed to return the index of the first appearance of an element, but the winning meaning-class does not include into the post-condition that the returned index is indeed the first appearance. 
Moreover, that formula was actually generated in another context and that led to the idea that the model was not able to detect that part of the intention in the docstring and formalize it in the same inference step. In general, this  phenomenon is known as the \textit{compositional gap}~\cite{DBLP:conf/emnlp/PressZMSSL23} in which models can correctly answer all sub-problems but not generate the overall solution. This, in turn, suggested that a structured docstring (i.e., NL pre-conditions followed by the NL post-conditions, both marked with an identification) could constitute a better subdomain for getting better performance in terms of the proposed distribution analysis. This could trigger, for instance, a re-engineering of the TM by adding a sentence close-ended classification task (pre-condition, post-condition, none) to help structuring stripped docstring as a preprocessing component added to the baseline TM.   

\begin{table}[h!]
    \caption{Nature of Misalignment for the Obtained Distributions.}
    \label{tab:error_breakdown_exp1}
\centering
   \resizebox{\columnwidth}{!}{%
 \begin{tabular}{|l|c|c|c|}
    \hline
    \textbf{Formalization Mistakes} & \textbf{Concentrated} & \textbf{Non-Concentrated} & \textbf{Total}\\
    \hline
    Weak post-condition        & 2  & 3 & 6 \\
    \hline
    Incorrect post-condition    & 1 & 2 & 4\\
    \hline
    Syntax error      & 1 & 2 & 3\\
    \hline
    Weak pre-condition          & 1 & 0 & 1\\
    \hline
    Strong pre-condition          & 1 & 0 & 0\\
    \hline
    \hline
    Total          & 6 & 7 & 14 \\
    \hline
    \end{tabular}}
    
\end{table}

In fact, in Table~\ref{tab:alignment_concentration_exp3} we show how the modified TM safely improves over the baseline. Vast majority of inputs lead to concentrated probability distributions and the concentration is on the right meaning class. Moreover, the number of misaligned and concentrated cases dropped from 6 to 2. Those two inputs leading to concentrated misalignment (indeed, not even a correct output generated), \textit{modify\_2d\_array} and \textit{on-line-max}, according to our proposal would be the focus of analysis for a new round of engineering efforts (out-of-scope). In fact, relevance of those inputs in focus can also be backed by the fact that in the baseline TM the resulting distributions were misaligned and non-concentrated (one without generation of the correct result). Moreover, \textit{modify\_2d\_array} was a paradigmatic case of non-concentration in the baseline TM (20 different meaning classes). On the other hand, both TMs were unable to generate one of the constraints required for the concept of maximum in \textit{on-line-max}. Through this guided analysis, we have identified natural language utterances where the model's assigned probabilities for formalizations are unsatisfactory.

\begin{table}[h!]
    \caption{Alignment and Concentration for the Modified TM.}
    \label{tab:alignment_concentration_exp3}
 \centering
   \resizebox{\columnwidth}{!}{%
    \begin{tabular}{|c|c|c|}
        \hline
        \textbf{Alignment/Concentration} & \textbf{Concentrated} & \textbf{Not Concentrated} \\
        \hline
        Aligned & 57 & 1 \\
        \hline
        Correct class gen. but misaligned distr. & 0 & 0 \\
        \hline
        Correct class not generated   & 2 & 2 \\ 
        \hline
    \end{tabular}}
\end{table}
\section{Related Work}
Surface form competition for multiple-choice tasks is introduced  in~\cite{DBLP:conf/emnlp/HoltzmanWSCZ21} and later discussed in~\cite{DBLP:conf/emnlp/WiegreffeFTCS23}. The  hallucination detection method of~\cite{DBLP:journals/nature/FarquharKKG24} works approximating entropy on meaning-classes, and equivalence checking for grouping results is done by means of LLMs given the diversity of benchmarked tasks. 
In~\cite{Clover}, transference model and similar notions of alignment and concentration are introduced in their analytical model to mathematically describe the hypothesis required to bound error in their program synthesis approach. In some sense, those works provide evidence on the potential usefulness of such concepts but do not elaborate on a framework to analyze and improve transference models.    

\section{Conclusions and Future Work}
We illustrate some principled engineering insights gained when probability distributions over meaning-classes are recovered from the stochastic behavior of transference models. This paper pinpoints at concentrated misalignment as key failure for ``debugging'' but inputs leading to non-concentrated distributions and their characterization might be also relevant. Sensitivity of distributions to input perturbations that are (or not) supposedly equivalent to the transformation under analysis could also be key to understand task-related limitations of the TM.  
Several RQs and experiments are necessary to claim generalizability and impact of the approach (e.g., In which extent tasks characteristics impact effectiveness of analysis?, In which extent knowing input distribution is key to analysis?, that is, When does improvement on a test set translate into generalized improvement?,  Which is the impact of decoding strategy in the analysis?, How robust the framework is to alternative practical distribution/equivalence approximation methods?,  etc.). There are also many possible conceptual and automation paths, including: approximation techniques for computing concentration,  the definition of a richer language to predicate on inputs and resulting distributions, and notions of compositionality of TMs. Also, we are aware that, for some tasks/transference models, a more structured domain of concepts might be needed (e.g., preference relation) and improvement definitions might need to accommodate potential trade-offs among lowering entropy and aligning distributions. 

Last but not least, it is foreseeable the  help of AI-based solutions to characterize misalignments and troublesome/adversarial inputs and even the assistance in decomposition, prompt-engineering and hyperparameter tuning.

\subsection{The case of Agentic AI}

A particular case of interest for safe AI is the analysis of agentic goal-pursuing LLM-enabled software. For such a case, we can naturally lift previously defined notions to a setting inspired by Markov Decision Processes~\cite{DBLP:books/wi/Puterman94}. 
In this vision, decisions of the MDP denote inputs from the environment (e.g., user-stated goals, invoked tools results, etc.). The LLM then provides the probability of transitioning to states that stand for  external action invocations or internal thoughts generated by the agent. More precisely, LLM can be regarded as implementing transference models that, given a representation of a history of interactions and thoughts, yields distributions on meaning classes each of them representing conceptually the same action invocation or same thoughts. 
In essence, an agentic LLM-powered software behaves as incrementally and reactively enacting a policy of action invocation and thoughts governed by the distributions over meaning-classes yielded by the LLM. Moreover, given an environmental policy or sequence of inputs/decisions, the MDP can be conceptually casted into a Markov Chain in which probability of a set of trajectories of interest can be theoretically measured.   
This formulation would be able to address unintended and competent misalignments. Firstly, the function from history to distributions on completions implemented by the LLM  is deemed deterministic  (there is no hidden scratchpad memory nor randomization available as mechanisms for implementing AI subversion strategies~\cite{DBLP:conf/icml/GreenblattSSR24}). Thus, one would be interested in strategies ``accidentally’’ played by the environment  in which trajectories that violate key safety constraints (e.g., no requesting human authorization) have a, higher than acceptable, chance to occur. 
Secondly, those inputs and feedback from the environment are not intentionally crafted (in the sense of jailbreaking) but they can be naturally expected according to some sensible operational profile governing external decisions.  Thirdly, they are not punctual or artificial:  they can be grouped into a measurable human-characterizable fault-inducing input class. Last but not least, violating trajectories look like competent behavior:  other safety (but less risky) constraints like environment preconditions for action execution are satisfied.


\begin{thebibliography}{10}
\interlinepenalty=10000
\providecommand{\url}[1]{\texttt{#1}}
\providecommand{\urlprefix}{URL }
\expandafter\ifx\csname urlstyle\endcsname\relax
  \providecommand{\doi}[1]{doi:\discretionary{}{}{}#1}\else
  \providecommand{\doi}{doi:\discretionary{}{}{}\begingroup \urlstyle{rm}\Url}\fi
\providecommand{\eprint}[2][]{\url{#2}}

\bibitem{repo-juan}
Juan Baldonado.
\newblock Source code for the experiments.
\newblock \texttt{\url{https://github.com/jmbuba/llm-semantic-perf}}, 2024.

\bibitem{braberman2024tasks}
V{\'{\i}}ctor~A. Braberman, Flavia Bonomo-Braberman, Yiannis Charalambous, Juan~G. Colonna, Lucas~C. Cordeiro, and Rosiane de~Freitas.
\newblock Tasks people prompt: A taxonomy of {LLM} downstream tasks in software verification and falsification approaches, 2024.
\newblock  \href{http://arxiv.org/abs/2404.09384}{{\scalebox{.9}[1.0]{\ttfamily arXiv:2404.09384}}}.

\bibitem{Few-Shot-Brown20}
Tom~B. Brown, Benjamin Mann, Nick Ryder, Melanie Subbiah, Jared Kaplan, Prafulla Dhariwal, et~al.
\newblock Language models are few-shot learners.
\newblock \textit{NeurIPS 2020}.

\bibitem{DBLP:journals/nature/FarquharKKG24}
Sebastian Farquhar, Jannik Kossen, Lorenz Kuhn, and Yarin Gal.
\newblock Detecting hallucinations in large language models using semantic entropy.
\newblock \textit{Nat.}, 630(8017):625--630, 2024.
\newblock \\ \href{https://doi.org/10.1038/S41586-024-07421-0}{{\ttfamily doi:10.1038/S41586-024-07421-0}}.

\bibitem{DBLP:conf/icml/GreenblattSSR24}
Ryan Greenblatt, Buck Shlegeris, Kshitij Sachan, and Fabien Roger.
\newblock {AI} control: Improving safety despite intentional subversion.
\newblock \textit{{ICML} 2024}.

\bibitem{hassan2024rethinking}
Ahmed~E. Hassan, Dayi Lin, Gopi~Krishnan Rajbahadur, Keheliya Gallaba, Filipe~Roseiro C{\^{o}}go, Boyuan Chen, et~al.
\newblock Rethinking software engineering in the era of foundation models: {A} curated catalogue of challenges in the development of trustworthy {FMware}.
\newblock \textit{{FSE} Comp. 2024}.
\newblock \\ \href{https://doi.org/10.1145/3663529.3663849}{{\ttfamily doi:10.1145/3663529.3663849}}.

\bibitem{DBLP:conf/emnlp/HoltzmanWSCZ21}
Ari Holtzman, Peter West, Vered Shwartz, Yejin Choi, and Luke Zettlemoyer.
\newblock Surface form competition: Why the highest probability answer isn't always right.
\newblock \textit{{EMNLP} 2021}.
\newblock \\ \href{https://doi.org/10.18653/V1/2021.EMNLP-MAIN.564}{{\ttfamily doi:10.18653/V1/2021.EMNLP-MAIN.564}}.

\bibitem{DBLP:journals/csur/JiLFYSXIBMF23}
Ziwei Ji, Nayeon Lee, Rita Frieske, Tiezheng Yu, Dan Su, Yan Xu, et~al.
\newblock Survey of hallucination in natural language generation.
\newblock \textit{{ACM} Comput. Surv.}, 55(12):248:1--248:38, 2023.
\newblock \\ \href{https://doi.org/10.1145/3571730}{{\ttfamily doi:10.1145/3571730}}.

\bibitem{DSPY}
Omar Khattab, Arnav Singhvi, Paridhi Maheshwari, Zhiyuan Zhang, Keshav Santhanam, Sri Vardhamanan, et~al.
\newblock {DSPy}: Compiling declarative language model calls into state-of-the-art pipelines.
\newblock \textit{{ICLR} 2024}.

\bibitem{dafny}
K.~Rustan~M. Leino.
\newblock Dafny: An automatic program verifier for functional correctness.
\newblock \textit{LPAR 2010}.
\newblock \\ \href{https://doi.org/10.1007/978-3-642-17511-4\_20}{{\ttfamily doi:10.1007/978-3-642-17511-4\_20}}.

\bibitem{RAG2020Patrick}
Patrick S.~H. Lewis, Ethan Perez, Aleksandra Piktus, Fabio Petroni, Vladimir Karpukhin, Naman Goyal, et~al.
\newblock Retrieval-augmented generation for knowledge-intensive {NLP} tasks.
\newblock \textit{NeurIPS 2020}.

\bibitem{DBLP:journals/toplas/LiskovW94}
Barbara Liskov and Jeannette~M. Wing.
\newblock A behavioral notion of subtyping.
\newblock \textit{{ACM} Trans. Program. Lang. Syst.}, 16(6):1811--1841, 1994.
\newblock \\ \href{https://doi.org/10.1145/197320.197383}{{\ttfamily doi:10.1145/197320.197383}}.

\bibitem{10.1162/tacl_a_00502}
Clara Meister, Gian Wiher, and Ryan Cotterell.
\newblock On decoding strategies for neural text generators.
\newblock \textit{Trans. Assoc. Comput. Linguistics}, 10:997--1012, 2022.
\newblock \\ \href{https://doi.org/10.1162/TACL\_A\_00502}{{\ttfamily doi:10.1162/TACL\_A\_00502}}.

\bibitem{openai2024gpt4technicalreport}
OpenAI, Josh Achiam, Steven Adler, Sandhini Agarwal, Lama Ahmad, Ilge Akkaya, et~al.
\newblock {GPT-4} technical report, 2024.
\newblock  \href{http://arxiv.org/abs/2303.08774}{{\scalebox{.9}[1.0]{\ttfamily arXiv:2303.08774}}}.

\bibitem{DBLP:conf/emnlp/PressZMSSL23}
Ofir Press, Muru Zhang, Sewon Min, Ludwig Schmidt, Noah~A. Smith, and Mike Lewis.
\newblock Measuring and narrowing the compositionality gap in language models.
\newblock \textit{{EMNLP} 2023}.
\newblock \\ \href{https://doi.org/10.18653/V1/2023.FINDINGS-EMNLP.378}{{\ttfamily doi:10.18653/V1/2023.FINDINGS-EMNLP.378}}.

\bibitem{DBLP:books/wi/Puterman94}
Martin~L. Puterman.
\newblock \textit{Markov Decision Processes: Discrete Stochastic Dynamic Programming}, Wiley, 1994.
\newblock \\ \href{https://doi.org/10.1002/9780470316887}{{\ttfamily doi:10.1002/9780470316887}}.

\bibitem{Clover}
Chuyue Sun, Ying Sheng, Oded Padon, and Clark~W. Barrett.
\newblock Clover: Closed-loop verifiable code generation.
\newblock \textit{{SAIV} 2024}.
\newblock \\ \href{https://doi.org/10.1007/978-3-031-65112-0\_7}{{\ttfamily doi:10.1007/978-3-031-65112-0\_7}}.

\bibitem{AutoformalizationSzegedy}
Christian Szegedy.
\newblock A promising path towards autoformalization and general artificial intelligence.
\newblock \textit{{CICM} 2020}.
\newblock \\ \href{https://doi.org/10.1007/978-3-030-53518-6\_1}{{\ttfamily doi:10.1007/978-3-030-53518-6\_1}}.

\bibitem{geminiteam2024geminifamilyhighlycapable}
Gemini Team, Rohan Anil, Sebastian Borgeaud, Jean-Baptiste Alayrac, Jiahui Yu, Radu Soricut, et~al.
\newblock Gemini: A family of highly capable multimodal models, 2024.
\newblock  \href{http://arxiv.org/abs/2312.11805}{{\scalebox{.9}[1.0]{\ttfamily arXiv:2312.11805}}}.

\bibitem{Self-consist}
Xuezhi Wang, Jason Wei, Dale Schuurmans, Quoc~V. Le, Ed~H. Chi, Sharan Narang, et~al.
\newblock Self-consistency improves chain of thought reasoning in language models.
\newblock \textit{{ICLR} 2023}.

\bibitem{CoT-Wei22}
Jason Wei, Xuezhi Wang, Dale Schuurmans, Maarten Bosma, Brian Ichter, Fei Xia, et~al.
\newblock Chain-of-thought prompting elicits reasoning in large language models.
\newblock \textit{NeurIPS 2022}.

\bibitem{DBLP:conf/emnlp/WiegreffeFTCS23}
Sarah Wiegreffe, Matthew Finlayson, Oyvind Tafjord, Peter Clark, and Ashish Sabharwal.
\newblock Increasing probability mass on answer choices does not always improve accuracy.
\newblock \textit{{EMNLP} 2023}.
\newblock \\ \href{https://doi.org/10.18653/V1/2023.EMNLP-MAIN.522}{{\ttfamily doi:10.18653/V1/2023.EMNLP-MAIN.522}}.

\bibitem{ReAct}
Shunyu Yao, Jeffrey Zhao, Dian Yu, Nan Du, Izhak Shafran, Karthik~R. Narasimhan, et~al.
\newblock {ReAct}: Synergizing reasoning and acting in language models.
\newblock \textit{{ICLR} 2023}.

\end{thebibliography}

\end{document}